\documentclass[conference]{IEEEtran}
\usepackage{caption}
\usepackage{subcaption}
\usepackage{color,soul}
\usepackage{amsthm} 
\theoremstyle{plain}

\theoremstyle{definition}

\theoremstyle{remark}

\usepackage{cite}
\usepackage{amsmath,amssymb,amsfonts}
\usepackage{algorithm}
\usepackage{algorithmic}
\usepackage{graphicx}
\usepackage{textcomp}
\usepackage{xcolor}
\usepackage{xpatch}
\usepackage{esint}
\usepackage[hidelinks]{hyperref}
\makeatletter
\ExplSyntaxOn
\cs_new:Npn \bibColoredItems #1#2
  {
    \clist_map_inline:nn {#2} { \cs_new:cpn {bib@colored@##1} {#1} } 
  }
\ExplSyntaxOff

\newcommand\bib@setcolor[1]{%
  \ifcsname bib@colored@#1\endcsname
    \expanded{\noexpand\color{\csname bib@colored@#1\endcsname}}%
  \else
    \normalcolor
  \fi
}

\IfPackageLoadedTF{hyperref}{\@tempswatrue}{\@tempswafalse}
\if@tempswa
  \xpatchcmd\@bibitem {\H@item}{\bib@setcolor{#1}\H@item}{}{\PatchFailed}
  \xpatchcmd\@lbibitem{\H@item}{\bib@setcolor{#2}\H@item}{}{\PatchFailed}
\else
  \xpatchcmd\@bibitem {\item}  {\bib@setcolor{#1}\item}  {}{\PatchFailed}
  \xpatchcmd\@lbibitem{\item}  {\bib@setcolor{#2}\item}  {}{\PatchFailed}
\fi
\makeatother

\usepackage{color}
\usepackage{bm}
\usepackage{booktabs}
\usepackage{cleveref}

\usepackage{geometry}
\ifCLASSINFOpdf

\else

\fi

\begin{document}
\title{
Wavenumber-Domain Near-Field Channel Estimation: Beyond the Fresnel Bound
\thanks{Xufeng Guo and Yuanbin Chen contributed equally to this work.
}

}

\author{{\large Xufeng~Guo\textsuperscript{1},~Yuanbin~Chen\textsuperscript{1},~Ying~Wang\textsuperscript{1},~Zhaocheng~Wang\textsuperscript{2},~and~Chau Yuen\textsuperscript{3}
}
\\
	{\normalsize \textsuperscript{1}State
Key Laboratory of Networking and Switching Technology,}\\
 {\normalsize Beijing University of Posts and Telecommunications, Beijing 100876, China} \\
	 {\normalsize \textsuperscript{2}Department of Electronic Engineering, Tsinghua University, Beijing 100084, China}   \\
{\normalsize \textsuperscript{3}School of Electrical and Electronic Engineering, Nanyang Technological University, Singapore 639798}
 
\vspace{-0.1cm}
}

\maketitle

\begin{abstract}
In the near-field context, the Fresnel approximation is typically employed to mathematically represent solvable functions of spherical waves. However, these efforts may fail to take into account the significant increase in the lower limit of the Fresnel approximation, known as the Fresnel distance. 
The lower bound of the Fresnel approximation imposes a constraint that becomes more pronounced as the array size grows. Beyond this constraint, the validity of the Fresnel approximation is broken. As a potential solution, the wavenumber-domain paradigm characterizes the spherical wave using a spectrum composed of a series of linear orthogonal bases. However, this approach falls short of covering the effects of the array geometry, especially when using Gaussian-mixed-model (GMM)-based von Mises-Fisher distributions to approximate all spectra.
To fill this gap, this paper introduces a novel wavenumber-domain ellipse fitting (WD-EF) method to tackle these challenges. 
Particularly, the channel is accurately estimated in the near-field region, by maximizing the closed-form likelihood function of the wavenumber-domain spectrum conditioned on the scatterers' geometric parameters.
Simulation results are provided to demonstrate the robustness of the proposed scheme against both the distance and angles of arrival.
\end{abstract}


\IEEEpeerreviewmaketitle

\section{Introduction}
 
Extremely large-scale antenna arrays (ELAA) play a crucial role in future sixth-generation (6G) communication systems, particularly in the high-frequency bands~\cite{HoloMag,chen}.
As the size of the array increases to the thousands and the carrier frequencies reach the terahertz level, the Rayleigh distance may be extended further to several hundred or even several thousand meters.
~\cite{TieruiGong,DLL-ULA}.
Given these circumstances, the spherical-wave impacts plays an important role in the design of near-field transmissions~\cite{DLL-ULA,DLL-UPA}.

According to the Huygens-Fresnel principle, the near-field channel response can be represented by the superposition of spherical waves from multiple scattering sources. 
Although a closed-form expression of these spherical waves can be obtained, the non-linear and non-uniform properties hinders the practical channel estimation designs. Particularly, 
the non-linearity may cause severe power leakage in the angular domain~\cite{guo-ICC24}; the non-uniform power of the spherical wave makes it intractable to design a proper codebook (or dictionary) to identify the source locations of the clusters in the propagation environment~\cite{DLL-ULA}.
The Fresnel approximation has been embraced as a straightforward technique to provide a tractable approximated representation of spherical waves~\cite{DLL-ULA, DLL-UPA, Guo-TPD}.
Based on this approximation, a customized polar-domain codebook is developed in~\cite{DLL-ULA}, along with an effective algorithm that estimates the distance between the transmitter and the scatterer/receiver. This approach has been further extended to the uniform planar array (UPA) scenario in~\cite{DLL-UPA}. Additionally, a geometry decomposition scheme based on the Fresnel approximation was proposed in~\cite{Guo-TPD} to exploit the symmetrical characteristics of the antenna elements, manifesting significantly reduced complexity from multiplicative to additive.

In examining the essence of the Fresnel approximation, the spherical wave is typically approximated by employing a paraboloid surface. The upper boundary of its effective range is determined by the Rayleigh distance~\cite{Yuanwei-Mag}.
When a scatterer is located within the Rayleigh distance, the Fraunhofer approximation, which treats the spherical wave as a planar wave in the far field, loses its effectiveness due to significant phase errors. 
In this case, the Fresnel approximation becomes essential. The upper bound of the Fresnel approximation is frequently discussed, while its lower limits, namely the Fresnel distance, is seldom addressed. 
Specifically, when the distance between the scatterer and the ELAA is too close below the Fresnel distance, the parabolic surface used in the Fresnel approximation might not exactly coincide with the spherical wave~\cite{Yuanwei-Mag}. 
Additionally, the fundamental differences between the reactive near-field channel below the Fresnel distance and the radiating near-field lie in the impact of the array geometry on the channel response~\cite{Yuanwei-Mag}. Thus, the closed-form spectrum is determined jointly by the scatterers' geometric parameters and the array geometry.

\begin{figure*}
  \begin{minipage}{0.58\linewidth}
      \centering
      \includegraphics[width=0.8\textwidth]{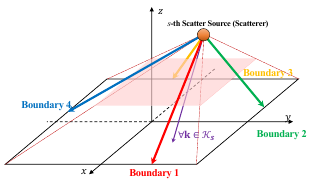}
      \subcaption{Physical boundaries of the $s$-th spherical wavefront.}
  \end{minipage}
  \hfill
  \begin{minipage}{0.365\linewidth}
      \centering
      \includegraphics[width=0.75\textwidth]{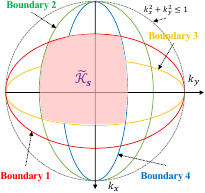}
      \subcaption{Corresponding elliptic boundaries of the wavenumber-domain spectrum area $\widetilde{\mathcal{K}}_s$.}
  \end{minipage}
  \caption{The wavenumber-domain spectrum in the reactive near field below the Fresnel distance.} \label{EllipticBoundary}
  \label{fig:enter-label}
  \vspace{-0.5cm}
\end{figure*}

The Fresnel distance increases significantly with higher carrier frequencies and larger antenna apertures. For instance, the Fresnel distance can extend to over one hundred meters with the array size of $512\times 512$ and the carrier frequency of $7\ {\rm GHz}$.
To effectively represent spherical waves within the Fresnel distance, rather than relying on approximation-based methods, we focus on the state-of-the-art wavenumber-domain representation~\cite{Fourier, guo-ICC24}. Specifically, this methodology employs the Weyl expansion to model spherical waves as a linear superposition of a finite number of planar waves, termed the Fourier harmonics~\cite{Fourier}. Theoretically, in the presence of the wavenumber-domain spectrum, the channel response generated by arbitrary scatterers can be characterized precisely in the free space~\cite{guo-ICC24}. 
However, the existing approaches in the wavenumber domain approximate the channel response as a von Mises-Fisher (vMF) distribution, which is actually a standard Gaussian mixture model (GMM)-based distribution~\cite{guo-ICC24, Fourier}. This falls short of accounting for the impact of array geometry on the wavefront, resulting in significant errors below the Fresnel distance. Therefore, there has been limited research focusing on the reactive near-field region below the Fresnel distance. Although we are able to use the wavenumber-domain spectrum to represent spherical waves, this representation is primarily limited to the simplest GMM-based vMF distributions. 

The contributions of this paper are summarized as follows:
\begin{itemize}
  \item An analytical expression of the wavenumber-domain spectrum is derived for the reactive near-field region below the Fresnel region. The spectrum boundaries in the UPA case are represented by the ellipses determined by scatterers' geometric parameters.
  \item Given that the boundaries encapsulate complete information about the scatterer's geometric parameters, it is logical to reconstruct them by determining the optimal solution for the likelihood function based on these parameters.
  This process is equivalent to fitting ellipses to the observed elliptic boundaries, i.e., our proposed wavenumber-domain ellipse fitting (WD-EF) scheme.

  \item Numerical results demonstrate that our proposed scheme significantly surpasses other benchmarks in estimation accuracy and robustness against variations in both distance and the angles of arrival (AoAs) under the Fresnel distance.
\end{itemize}

\section{System Model}

\begin{figure*}
  \begin{minipage}[t]{0.355\linewidth}
      \centering
      \vspace{0pt} 
      \includegraphics[width=0.75\textwidth]{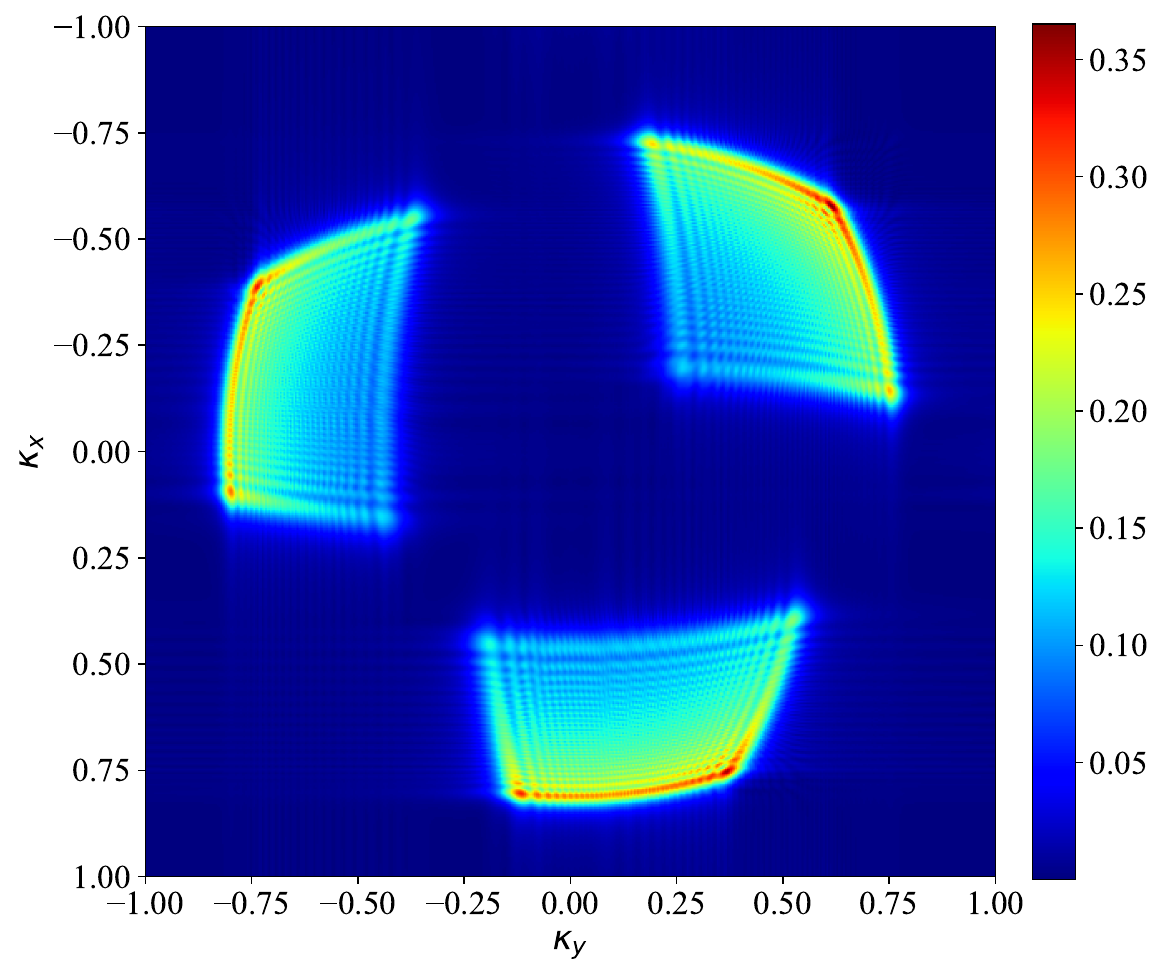}
      \subcaption{The wavenumber-domain spectrum.}
  \end{minipage}
  \hfill 
  \begin{minipage}[t]{0.315\linewidth}
      \centering
      \vspace{0pt}
      \includegraphics[width=0.75\textwidth]{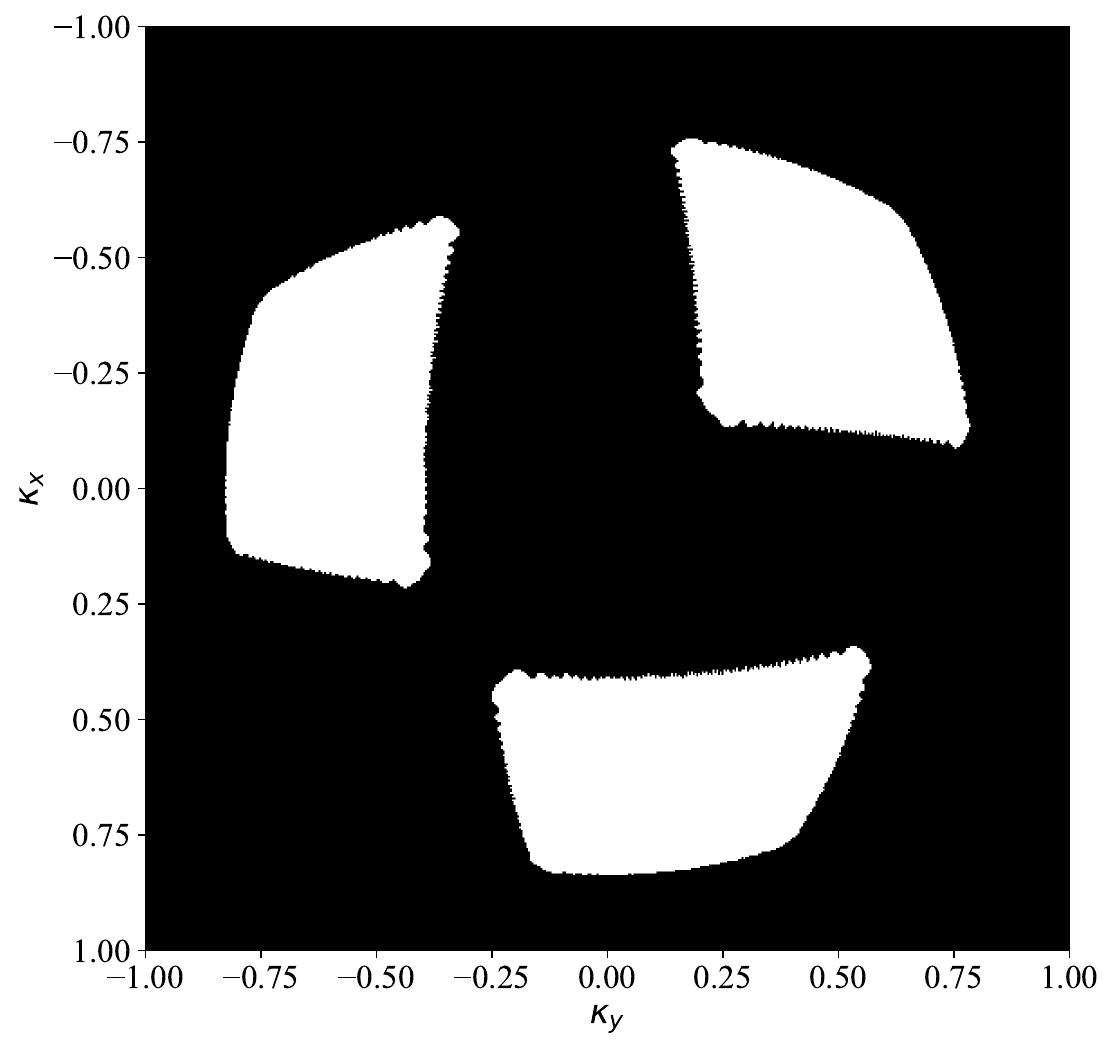}
      \subcaption{The non-zero indicators filtered by power threshold.}
  \end{minipage}
  \hfill 
  \begin{minipage}[t]{0.305\linewidth}
      \centering
      \vspace{0pt}
      \includegraphics[width=0.75\textwidth]{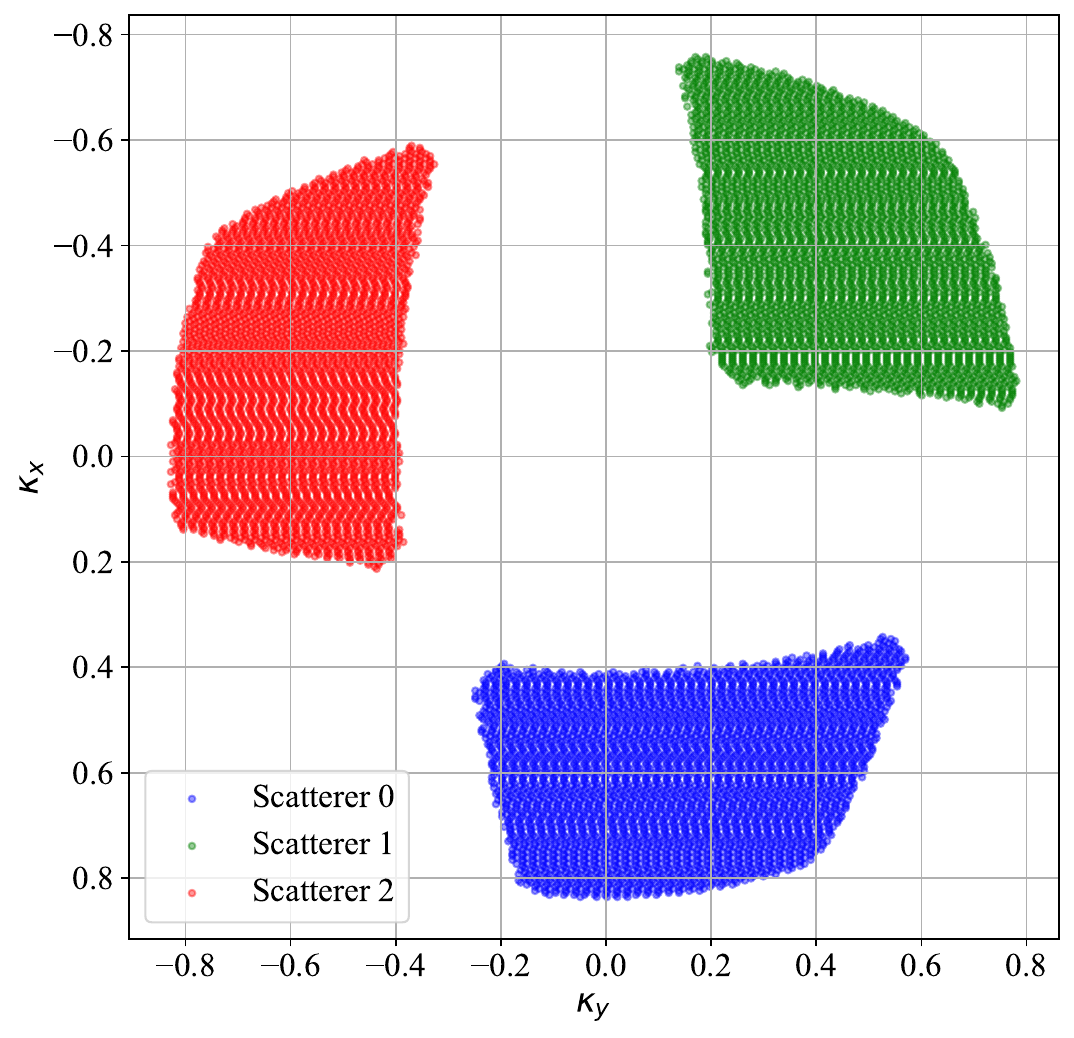}
      \subcaption{The clustering results of the non-zero indicators using GMM~\cite{guo-ICC24}.}
  \end{minipage}
  \caption{Pre-process before the execution of the proposed WD-EF algorithm, Fresnel distance is $100\ {\rm m}$, scatterer distance $r_s = 15\ {\rm m}$, scatterer number $S  = 3$.}
  \label{fig_WDEF}
  \vspace{-0.5cm}
\end{figure*}
We consider an uplink multiple-input-multiple-output (MIMO) communication system, with the receiver equipped with an extremely large-scale antenna array (ELAA) and arranged as the uniform planar array (UPA).
The UPA is aligned with the $xoy$ plane with $N = N_x \times N_y$ antenna elements, where $N_x$ and $N_y$ denote the number of antennas along the $x$ and $y$ axis, respectively.
Therefore, the array aperture can be expressed as $L = {\rm norm}([L_x, L_y])$, where $L_x=(N_x - 1) \delta$, $L_y=(N_y - 1) \delta$.
The electromagnetic (EM) field is assumed to be time-invariant within the channel's coherence time and determined by $S$ scatterers within the EM field, each corresponding to a propagation path and wavefront at the MIMO side.
The Cartesian coordinate of the $s$-th scatterer is denoted by
\begin{equation}\label{est}\small
  {\bf r}_s = \left[
    \begin{array}{c}
      r_{s}^x \\
      r_{s}^y \\
      r_{s}^z
    \end{array}
  \right] = r_s  {\bf k}_s
  =r_s  \left[
    \begin{array}{c}
      k_s^x \\
      k_s^y \\
      k_s^z
    \end{array}
  \right]
  = r_s  \left[
    \begin{array}{c}
      \sin \theta_s \cos \phi_s \\
      \sin \theta_s \sin \phi_s \\
      \cos \theta_s
    \end{array}
  \right],
\end{equation}
where ${\bf k}_s$ is the unit wavenumber vector that indicates the direction of the $s$-th scatterer. $(r_s, \theta_s, \phi_s)$ denotes the spherical coordinate of the $s$-th scatterer with $r_s$, $\theta_s$, and $\phi_s$ being the distance, the elevation angle, and the azimuth angle, respectively.

The steering matrix ${\bf A} \in \mathbb{C}^{N_x\times N_y\times S}$ generated by the $S$ scatterers indexed by $\mathcal{S} = \{{0,\dots, S-1}\}$ is given by
\begin{equation}\label{eq:steering_matrix}
  \left[{\bf A}\right]_{n_x, n_y, s} =
  g_s d_{n_x, n_y, s}^{-1} \exp \left\{
    -j k_c d_{n_x, n_y, s}
  \right\},
\end{equation}
where $k_c = 2\pi/\lambda$ is the carrier wavenumber, $g_s$ is the complex gain of the $s$-th scatterer, and $d_{n_x, n_y, s}$ is the distance between the $s$-th scatterer and the $(n_x, n_y)$-th antenna element. 
  Note that when modeling the near-field channel, the steering matrix in~\eqref{eq:steering_matrix} encapsulates the non-uniform characteristics where the complex channel gain with respect to (w.r.t.) each antenna element is different, i.e., $g_sd_{n_x, n_y,s}^{-1}$, due to different distances between the scatterers and the antenna elements.
Besides the non-uniform property, another unfavorable characteristic of the near-field channel is the non-linear phase sequence w.r.t. the index of the antenna element.
Although the Fresnel approximation is used to make a second-order approximation of the spherical phase difference ${d}_{n_x, n_y, s}$, it fails to eliminate phase errors between the perfect spherical wave and the approximated paraboloid. This modeling error becomes significantly worse below the Fresnel distance~\cite{Yuanwei-Mag}, defined as $0.5\left(D^3 / \lambda\right)^{0.5}$. Below this bound, the Fresnel approximation becomes invalid.



\section{The Proposed Wavenumber-Domain Ellipse Fitting (WD-EF) Scheme}

Given the nonlinear and non-uniform properties of spherical waves shown in~\eqref{eq:steering_matrix}, we focus on the use of the wavenumber-domain channel representation~\cite{Fourier,guo-ICC24}. This method provides a spatially stationary equivalent to the original spherical wavefronts~\cite{Fourier}. Based on this representation, we derive the closed-form expressions for both the amplitude and the boundaries of the wavenumber-domain spectrum in Sec.\ref{amp} and Sec.\ref{BoundarySection}, respectively.
An example of the wavenumber-domain spectrum under the Fresnel distance can be found in Fig.~\ref{fig_WDEF}(a).
As it transpires, the spectrum of the scatterer is not perfect round shape as assumed in the GMM-based method; instead, it is constrained in the square-like region determined by four boundaries.
Since the spectrum boundaries encapsulate the complete information regarding the scatterer's geometric parameters, we propose a novel WD-EF algorithm in Sec.~\ref{AlgSec}. This algorithm recovers the channel characteristics by fitting the boundaries with the optimal scatterer-parameter-determined ellipses.

\subsection{Wavenumber-Domain Spectrum of Spherical Waves}\label{amp}
We first observe the array response generated by the $s$-th scatterer $a_s(x,y)$ at a sampling point on the MIMO surface with the coordinate $(x, y, 0)$ according to the Weyl expansion as follows
\begin{equation}\label{weyl}
\small
\begin{aligned}  
&a_s(x,y) 
=\frac{g_s}{j2\pi} \oiint_{{\bf k}\in \mathcal{K}_s} k_z^{-1}\exp\{-jk_c{\bf k}\cdot ({\bf p}-{\bf r}_s)\}
\\
&=\frac{jg_s}{2\pi} \oiint_{{\bf \kappa} \in \mathcal{\bar K}_s}  \kappa_z^{-1} \mathcal{P}_s(-\kappa_x,- \kappa_y) \exp \{      -j {k_c} (\kappa_x x + \kappa_y y)     \}\ {\rm d} \kappa_x {\rm d} \kappa_y,  \end{aligned}
\end{equation}
where $\mathcal{P}_s (\cdot, \cdot)$ represents a phase shift factor with a modulus of one, given by
\begin{equation}\label{PhaseTerm}
    \mathcal{P}_s(-\kappa_x, -\kappa_y)) = \exp\left\{j(k_x r_s^x+ k_yr_s^y + k_z r_s^z)\right\}.
\end{equation}
${\bf k} \triangleq [k_x, k_y, k_z]^T$ denotes the unit wavenumber vector that indicates the wave propagation direction.
$\boldsymbol{\kappa} \triangleq - {\bf k} = [\kappa_x,\kappa_y,\kappa_z]^T$ is defined as the opposite direction of the propagation direction, indicating the incident direction.
Both ${\bf k}$ and $\boldsymbol{\kappa}$ satisfy $||{\bf k}|| = ||\boldsymbol{\kappa}|| = 1$.
The term $\mathcal{K}_s$ represents the integration range for ${\bf k}$, constituting a three-dimensional linear space. This space encompasses the region of wave propagation directions for all non-zero responses produced by the $s$-th scatterer.
Conversely, $\widetilde{\mathcal{K}}_s$ is similar to $\mathcal{K}_s$ but represents the region of wave incident directions with opposite orientation.

If we consider the integration over the continuous wavenumber domain, i.e., $\oiint_{{\boldsymbol \kappa}\in \widetilde{\mathcal{K}}_s}$, this can be understood as the superposition of innumerable small waves, each with an incident direction given by ${\boldsymbol \kappa}$. 
It should be highlighted that in~\eqref{weyl}, only the component $\exp \{-j (k_x x + k_y y)\}, \forall {\bf k}\in \mathcal{K}_s$ indicates the wave propagation directions. Therefore, it is safe to say that each infinitesimal wave corresponding to the incident direction $\boldsymbol{\kappa} = -{\bf k}$ within the integration region $\widetilde{\mathcal{K}}_s$ is a planar wave given by
$
\exp\{-j k_c (\kappa_x x + \kappa_y y)\}
$
with the amplitude of
\begin{equation}
\begin{aligned}
    \mathcal{A}_s (\kappa_x, \kappa_y) &= |\kappa_z^{-1} \mathcal{P}_s(-\kappa_x, -\kappa_y))| 
    \\
    &= (1 - (\kappa^2_x + \kappa^2_y))^{-1/2}, \quad\forall (\kappa_x, \kappa_y) \in \mathcal{\widetilde{K}}_s.
\end{aligned}
\end{equation}
where $\mathcal{A}_s (\kappa_x, \kappa_y)$ denotes the corresponding wavenumber-domain spectrum of the $s$-th scatterer.


\subsection{Boundary Derivations}\label{BoundarySection}

Upon reaching the antenna array, the spherical wave generated by the $s$-th scatterer retains only a portion of the wavefront. This preserved wavefront is determined by the scatterer's position, denoted as ${\bf r}_s$, and the geometry of the receiving antenna array. 
As illustrated in Fig.~\ref{EllipticBoundary}(a), the boundaries can be categorized into four kinds, determined by the four edges of the UPA panel.
Moreover, the boundaries of the corresponding wavenumber-domain spectrum are characterized by perfect elliptic shapes, as shown in~Fig.~\ref{EllipticBoundary}(b). This geometric interpretation provides a handle for establishing the relationship between the boundaries and the estimated position of the scatterer, ${\bf r}_s$, as will be discussed as follows.

\subsubsection{\textbf{Boundary 1\&3}}

We first solve for the expression of the curve \textbf{Boundary~1} in the wavenumber domain, denoted as $\mathcal{C}_{1}$. 
Based on the direction of the incoming wave and the geometric characteristics of the UPA, it is observed that each point on \textbf{Boundary~1} in the wavenumber domain should maintain a parallel orientation to the propagation direction of the wave from the $s$-th scatterer to the boundary points of the UPA, specifically at $[L_x/2, y, 0]^T$, where $y \in [-L_y/2, L_y/2]$.
Therefore, each point $(\kappa_x, \kappa_y)$ on $\mathcal{C}_{1}$ in the wavenumber domain satisfies the following constraint
\begin{equation}\label{C1C}\small
    \boldsymbol{\kappa} = \left[\begin{array}{ccc}
             \kappa_x\\
             \kappa_y\\
            \kappa_z
        \end{array}
    \right] = {\rm norm}\left(-{\bf r}_s + \left[\begin{array}{c}
0.5 L_x \\
y \\
0
\end{array}\right]\right),\ \forall (\kappa_x, \kappa_y) \in \mathcal{C}^1_s,
\end{equation}
where $\kappa_z = \sqrt{1-\kappa_x^2 - \kappa_x^2}$ is the wavenumber-domain component along the $z$-axis, $L_x$ is the array aperture along the $x$-axis. 
From~\eqref{C1C}, we can obtain two separated constrains for $\kappa_x, \kappa_y$, respectively as follows
\begin{equation}\label{C1C2}\small
    {\kappa}_x=\frac{r^x_s-0.5 L_x}{\left\|-{\bf r}_s+\left[\begin{array}{c}
0.5 L_x \\
y \\
0
\end{array}\right]\right\|}, \quad {\kappa}_y=\frac{r^y_s-y}{\left\|-{\bf r}_s+\left[\begin{array}{c}
0.5 L_x \\
y \\
0
\end{array}\right]\right\|}.
\end{equation}
Upon dividing the sub equations in~\eqref{C1C2}, we can get the substitution expression for the variable $y$
\begin{subequations}\label{C1C3}
\begin{align}
    \frac{\kappa_y}{\kappa_x}&=\frac{r_s^y-y}{r_s^x-0.5 L_x},\\
    & \xrightarrow{\text { yields }} y=\frac{\kappa_y}{\kappa_x}\left(-r_s^x+0.5 L_x\right)+r_s^y.\label{subs_y}
\end{align}
\end{subequations}
By expressing $y$ in~\eqref{C1C2} as the form specified in~\eqref{subs_y}, we can derive a curve expression that solely depends on the variables ${\kappa_x, \kappa_y}$
\begin{equation}\label{C1}
    \mathcal{C}^1_s\left(\kappa_x, \kappa_y\right)\triangleq\left\{
    \begin{split}
        &\frac{\left(0.5 L_x-r_s^x\right)^2+\left(r_s^z\right)^2}{\left(-r_s^x+0.5 L_x\right)^2} {\kappa}_x^2+{\kappa}_y^2=1;\\
    &\operatorname{sign}\left({\kappa}_x\right)=\operatorname{sign}\left(r_s^x-0.5 L_x\right)
    \end{split}\right\},
\end{equation}
where ${\rm sign}\left(\cdot \right)$ is the sign function.

Similar to~\eqref{C1C}, we can writing the geometric constrain of \textbf{Boundary 3}, i.e., $\forall (\kappa_x, \kappa_y) \in \mathcal{C}^3_s$ as
\begin{equation}\label{C3C}
    \boldsymbol{\kappa} = {\rm norm}(-{\bf r}_s + [
-0.5 L_x, y
, 0]^T),\ \forall (\kappa_x, \kappa_y) \in \mathcal{C}^3_s.
\end{equation}
Then upon the same substitution process described in~\eqref{C1C2} and~\eqref{C1C3}, we can obtain the elliptic expression of $\mathcal{C}^3_s$ as
\begin{equation}\label{C3}
    \mathcal{C}^3_s\left(\kappa_x, \kappa_y\right)\triangleq\left\{
    \begin{split}
        &\frac{\left(0.5 L_x+r_s^x\right)^2+\left(r_s^z\right)^2}{\left(r_s^x+0.5 L_x\right)^2} {\kappa}_x^2+{\kappa}_y^2=1;\\
    &\operatorname{sign}\left({\kappa}_x\right)=\operatorname{sign}\left(r_s^x+0.5 L_x\right)
    \end{split}\right\}.
\end{equation}

\subsubsection{\textbf{Boundary 2\&4}}
Following the same reasoning as for \textbf{Boundary 1\&3}, since \textbf{Boundary 2\&4} are the y-axis counterparts, there exists a perfect reciprocity between them. Consequently, the derivation for \textbf{Boundary 2\&4} simply involves substituting the $x$ subscript with a $y$ subscript in equations~\eqref{C1} and~\eqref{C3}. Therefore, $\mathcal{C}^2_s$ and $\mathcal{C}^4_s$ can be expressed as
\begin{equation}\label{C2}
    \mathcal{C}^2_s\left(\kappa_x, \kappa_y\right)\triangleq\left\{
    \begin{split}
        &\frac{\left(0.5 L_y-r_s^y\right)^2+\left(r_s^z\right)^2}{\left(-r_s^y+0.5 L_y\right)^2} {\kappa}_y^2+{\kappa}_x^2=1;\\
    &\operatorname{sign}\left({\kappa}_y\right)=\operatorname{sign}\left(r_s^y-0.5 L_y\right)
    \end{split}\right\},
\end{equation}
and 
\begin{equation}\label{C4}
    \mathcal{C}^4_s\left(\kappa_x, \kappa_y\right)\triangleq\left\{
    \begin{split}
        &\frac{\left(0.5 L_y+r_s^y\right)^2+\left(r_s^z\right)^2}{\left(r_s^y+0.5 L_x\right)^2} {\kappa}_y^2+{\kappa}_x^2=1;\\
    &\operatorname{sign}\left({\kappa}_y\right)=\operatorname{sign}\left(r_s^y+0.5 L_y\right)
    \end{split}\right\},
\end{equation}
respectively.

\begin{figure*}[t!]
  \begin{minipage}[t]{0.33\linewidth}
      \centering
      \includegraphics[width=0.8\textwidth]{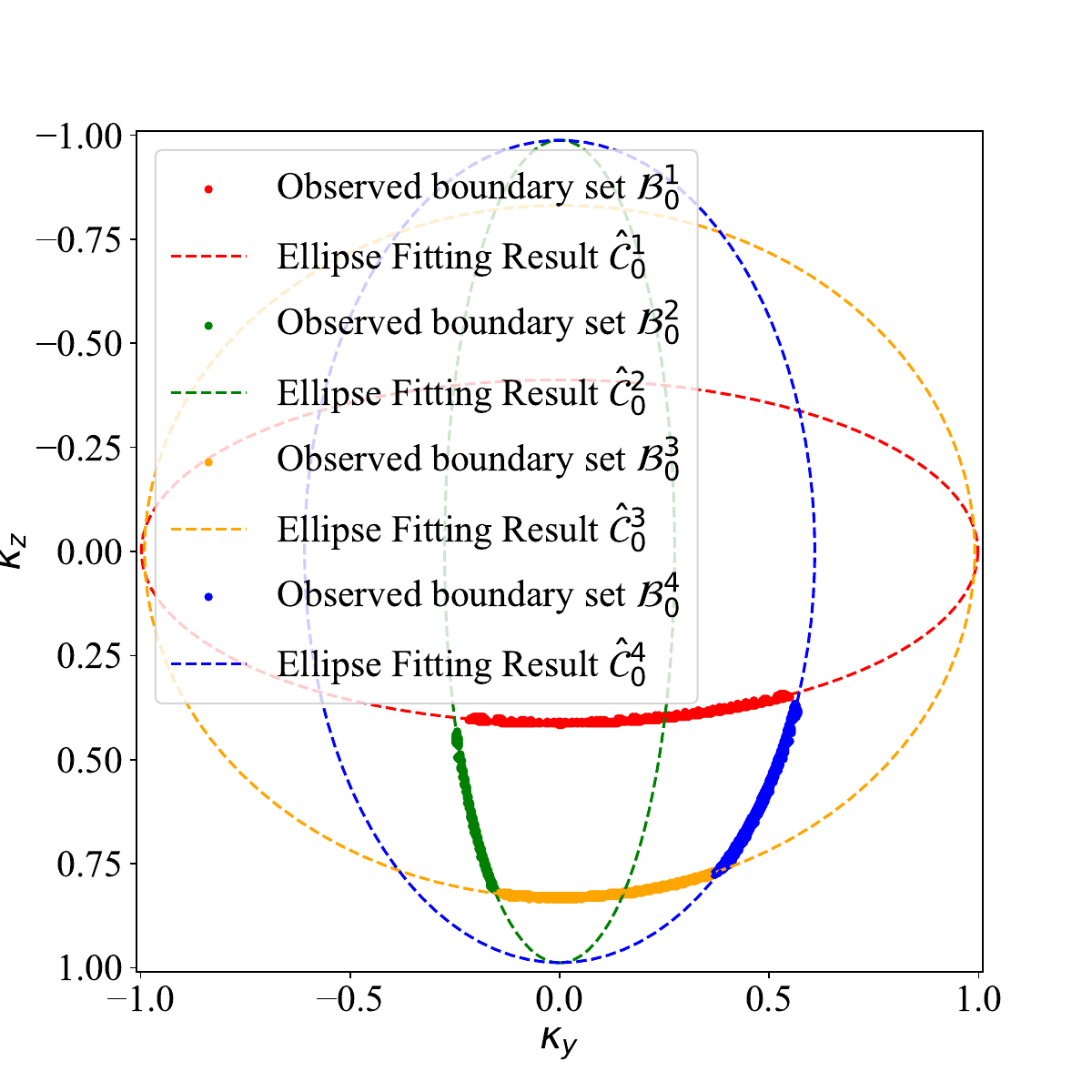}
      \subcaption{$(s = 0)$-th scatterer.}
  \end{minipage}
  \begin{minipage}[t]{0.33\linewidth}
      \centering
      \includegraphics[width=0.8\textwidth]{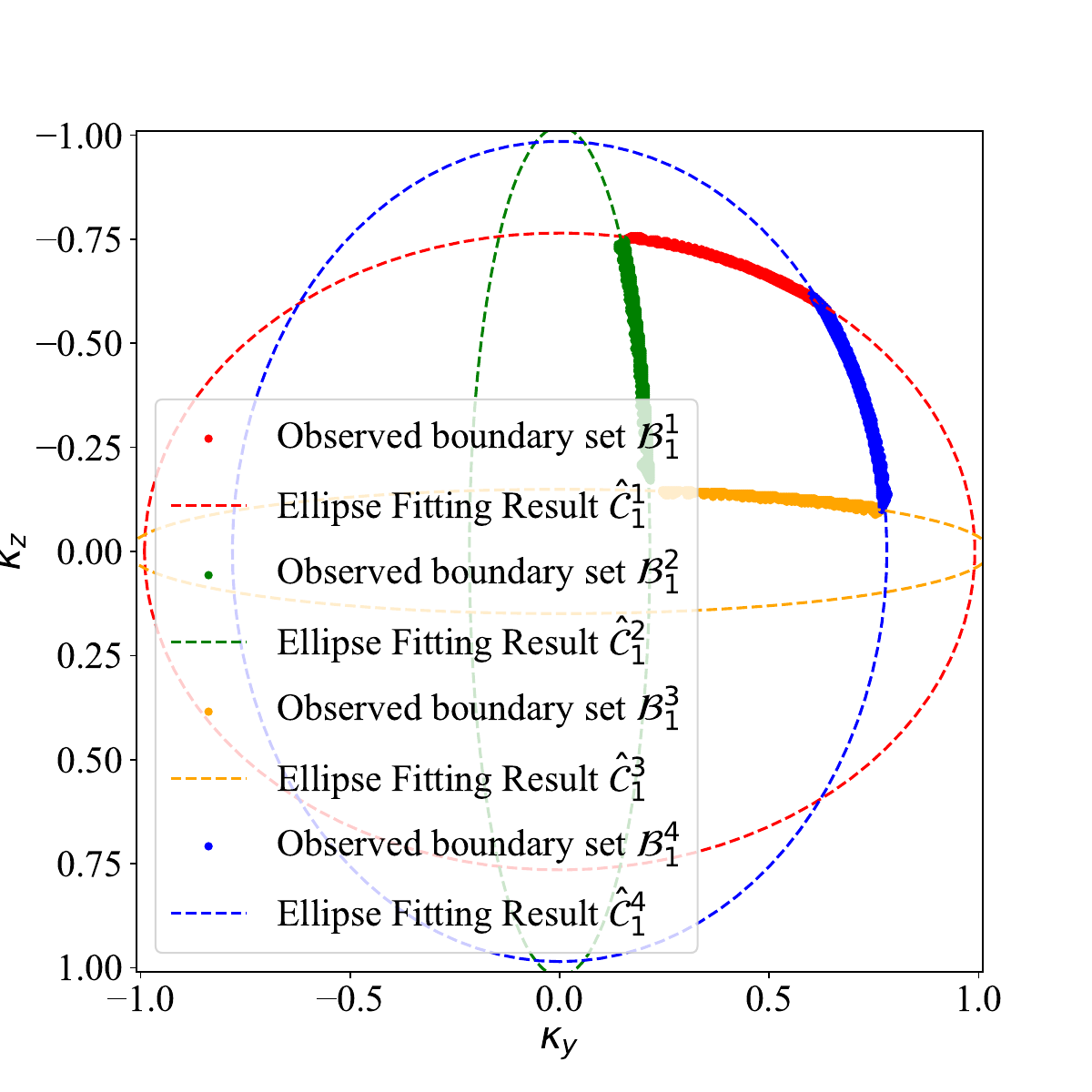}
      \subcaption{$(s = 1)$-th scatterer.}
  \end{minipage}
  \begin{minipage}[t]{0.33\linewidth}
      \centering
      \includegraphics[width=0.8\textwidth]{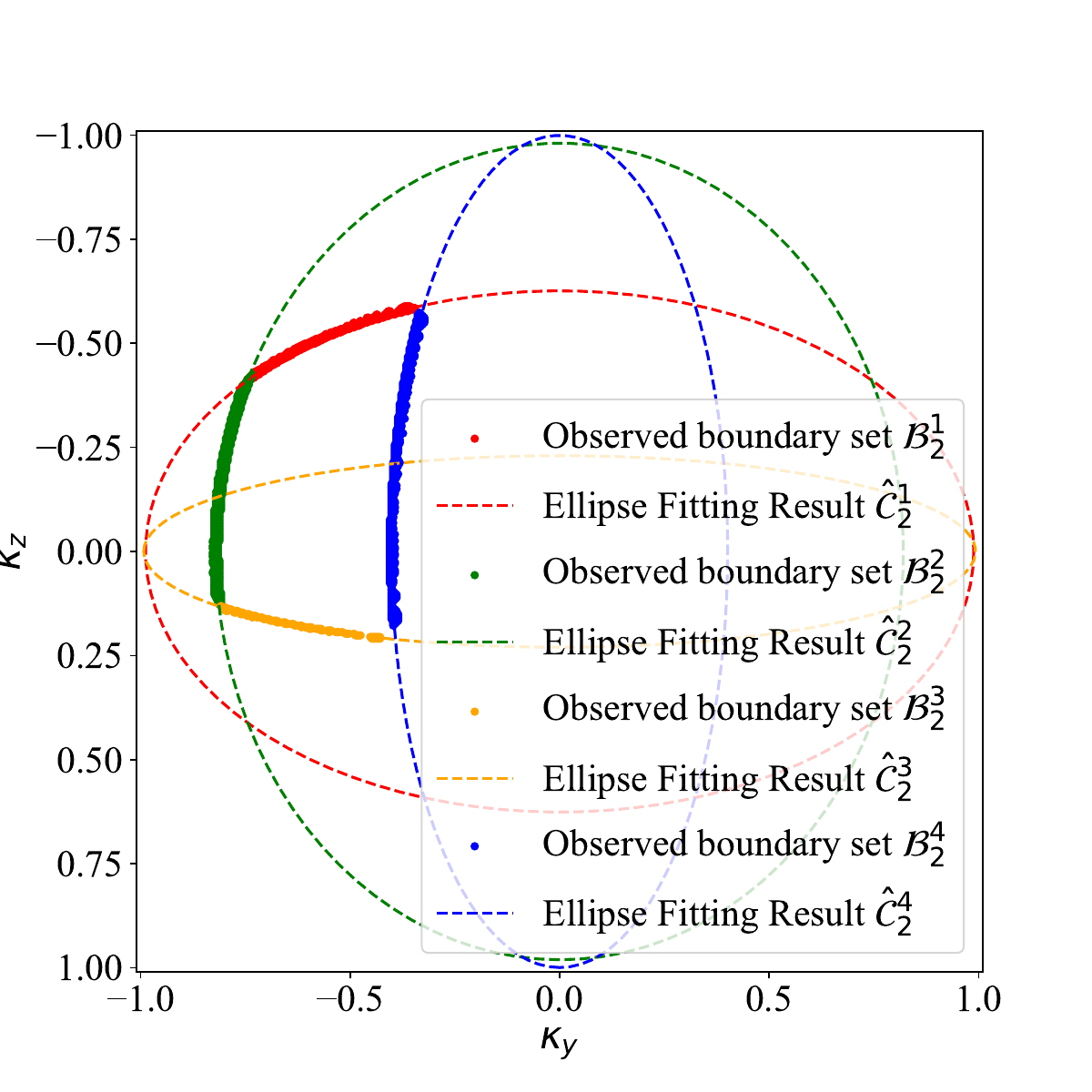}
      \subcaption{$(s = 2)$-th scatterer.}
  \end{minipage}
  \caption{The example of ellipse fitting, $S=3$, $r_s = 15\ {\rm m}, \theta_s = 45^\circ, \forall s\in \mathcal{S}$.}
  \label{fig_Algorithm}
  \vspace{-0.5cm}
\end{figure*}

\subsection{Parametric Estimation By Ellipse Fitting}\label{AlgSec}
\subsubsection{\textbf{Boundary Extraction}}
{\color{black}
The purpose of our proposed WD-EF algorithm is to accurately determine all geometric parameter estimates of $\{r^x_s, r^y_s,r^z_s\}$ (or their spherical counterparts $\{r_s, \theta_s, \phi_s\}$ in~\eqref{est}) based on the received wavenumber-domain power spectrum, as illustrated in Fig.~\ref{fig_WDEF}(a). Upon the observed wavenumber-domain spectrum, a simple power threshold technique is employed to extract the non-zero elements, as depicted in Fig.~\ref{fig_WDEF}(b). 
Following this, a classic clustering algorithm, specifically the Gaussian mixed model (GMM), is utilized to segregate different wavefronts, each associated with a distinct scatterer, as shown in Fig.~\ref{fig_WDEF}(c)\footnote{The optimal number of scatterers is determined by minimizing the Silhouette score~\cite{SilhouetteCoefficient}.}.}

{\color{black}
Upon the pre-processing illustrated in Fig.~\ref{fig_WDEF}, we have obtained each scatterer's distinct wavenumber-domain spectrum region, i.e., $\widetilde{\mathcal{K}}_s$.
Note that we have already derived the precise expressions for the boundaries of $\widetilde{\mathcal{K}}_s$, as determined by the geometric parameters ${r^x_s, r^y_s, r^z_s}$, in Sec.~\ref{BoundarySection}. Moving forward, we can extract the four boundaries\footnote{Note that \eqref{BoundaryExtraction} is the definition of the boundary extraction process. The actual boundary extraction process is based on the edge detection algorithm, such as the Canny algorithm~\cite{Canny}.} of the observed $\widetilde{\mathcal{K}}_s$
\begin{equation}\label{BoundaryExtraction}
    \mathcal{B}^j_s = \left\{
      (\kappa_x, \kappa_y) \in \widetilde{\mathcal{K}}_s: || \mathcal{C}^j_s (\kappa_x, \kappa_y) || \leq \epsilon, j \in \{1, 2, 3, 4\}
    \right\},
\end{equation}
where $\epsilon$ is a small positive number that determines how close should the observed boundary be to the theoretical boundary.
}

\subsubsection{\textbf{Ellipse Fitting}}
{\color{black}
Next, we aim to fit the observed boundary sample points $\mathcal{B}^j_s$ with the corresponding elliptic shapes $\mathcal{C}^j_s, s\in \mathcal{S}, j\in \{1,2,3,4\}$. 
We adopt the direct ellipse fitting algorithm proposed in~\cite{DirectLSEllipseFitting} to achieve this goal. 
Specifically, we design the observation matrix $\mathbf{D}^j_s\in \mathbb{R}^{|\mathcal{B}^j_s|\times 6}$ as
\begin{equation}\label{ObservationMatrix}
    \mathbf{D}^j_s
     = \left[
      \begin{array}{cccccc}
        \vdots & \vdots & \vdots & \vdots & \vdots & \vdots \\
        x_b^2 & x_b y_b & y_b^2 & x_b & y_b & 1 \\
        \vdots & \vdots & \vdots & \vdots & \vdots & \vdots
      \end{array}
    \right], \forall (x_b, y_b) \in \mathcal{B}^j_s.
\end{equation}
Then we define a constraint matrix $\mathbf{C}$ as:
\begin{equation}
  \mathbf{C}=\left[\begin{array}{cccccc}
    0 & 0 & 2 & 0 & 0 & 0 \\
    0 & -1 & 0 & 0 & 0 & 0 \\
    2 & 0 & 0 & 0 & 0 & 0 \\
    0 & 0 & 0 & 0 & 0 & 0 \\
    0 & 0 & 0 & 0 & 0 & 0 \\
    0 & 0 & 0 & 0 & 0 & 0
    \end{array}\right].
\end{equation}
By satisfying the following constraints:
  \begin{align}\label{EllipseFitting}
    \mathbf{S}_s^j \boldsymbol{\xi}_s^j  =\lambda \mathbf{C} \boldsymbol{\xi}_s^j, 
    (\boldsymbol{\xi}_s^j)^{\mathrm{T}} \mathbf{C} \boldsymbol{\xi}_s^j  =1,
    \end{align}
where $\mathbf{S}_s^j = \mathbf{D}^j_s (\mathbf{D}^j_s)^{\mathrm{T}}$, $\lambda$ is the eigenvalue, $\boldsymbol{\xi}_s^j = [a_s^j, b_s^j, c_s^j, d_s^j, e_s^j, f_s^j]^{\mathrm{T}}$ is the ellipse parameter vector so that the standard ellipse equation can be expressed as
\begin{equation}
  a_s^j x^2 + b_s^j x y + c_s^j y^2 + d_s^j x + e_s^j y + f_s^j = 0.
\end{equation}
There are up to six solutions of $\boldsymbol{\xi}_s^j$ in~\eqref{EllipseFitting}, and we choose the one with the smallest positive eigenvalue as the final solution.
Then, recall the expressions of~\eqref{C1}, \eqref{C3}, we can obtain the following elliptic estimation results for {\textbf{Boundary 1\&3}}
\begin{equation}\label{EFest13}
    (a^1_s  - 1) \cdot (
    -\hat{r}_s^x + 0.5 L_x
  )^2
  = \hat{r_s^z}^2 = 
  (a^3_s  - 1) \cdot (
    \hat{r}_s^x + 0.5 L_x
  )^2,
\end{equation}
which yields a quadratic equation for $\hat{r}_s^x$, given by
  $\alpha (r_s^x)^2 + \beta r_s^x + 0.25 \alpha L_x^2 = 0$,
where $\alpha = 1- \frac{a^1_s - 1}{a^3_s - 1}$ and $\beta = 1 + \frac{a^1_s - 1}{a^3_s - 1}$. Thus, the estimated $r_s^x$ can be obtained by solving the above quadratic equation, given by
\begin{equation}\label{rxEst}
  \begin{aligned}
    \hat{r}_s^x & = \frac{-\beta L_x \pm L_x \sqrt{\beta^2-\alpha^2}}{2 \alpha},
  \\
  {\rm sign}(\hat{r}_s^x) &= {\rm sign}(\hat{r}_s^x - 0.5 L_x).
  \end{aligned}
\end{equation}
By substituting $\hat{r}_s^x$ into~\eqref{EFest13}, we can obtain the estimated $r_s^z$ as
\begin{equation}\label{rzEst}
  \hat{r}_s^z = \sqrt{(a^1_s - 1) \cdot (
    -\hat{r}_s^x + 0.5 L_x
  )^2}.
\end{equation}
Using the reciprocal logic, we can obtain the estimated $r_s^y$ for \textbf{Boundary 2\&4} as:
\begin{equation}\label{ryEst}
  \hat{r}_s^y = \sqrt{(a^2_s - 1) \cdot (
    -\hat{r}_s^y + 0.5 L_y
  )^2}.
\end{equation}
Upon the estimation of $\{r^x_s, r^y_s, r^z_s\}$, we can further derive the spherical coordinates $\{r_s, \theta_s, \phi_s\}$ using the conversion relations in~\eqref{est}.
The overall WD-EF algorithm is summarized in \textbf{Algorithm~\ref{WD-EF}}.}

\begin{figure*}
  \centering
  \begin{minipage}{0.48\linewidth}      \centering\includegraphics[width=0.8\textwidth]{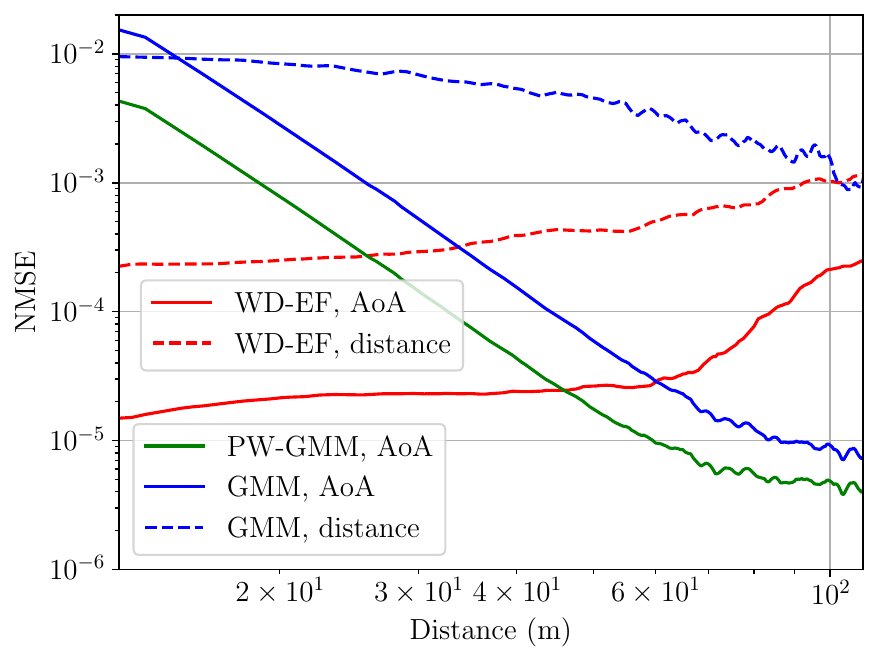}
      \subcaption{NMSE versus distance, $\theta_s = 45^\circ$, $\forall s\in \mathcal{S}$.}
  \end{minipage}
  \begin{minipage}{0.48\linewidth}
      \centering\includegraphics[width=0.8\textwidth]{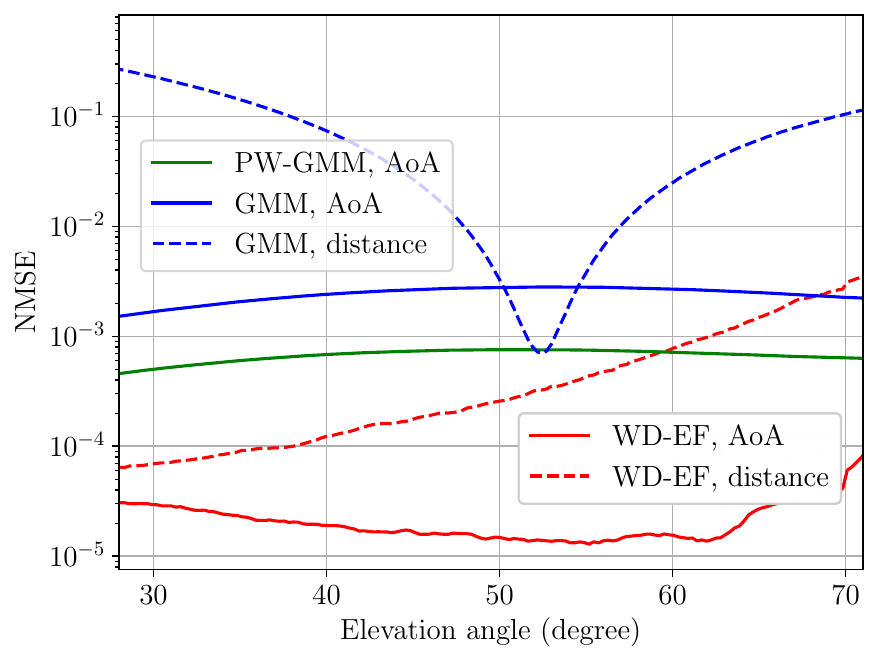}
      \subcaption{NMSE versus elevation, $r_s = 20\ {\rm m}$, $\forall s \in \mathcal{S}$.}
  \end{minipage}
  \caption{Simulation results, Fresnel distance is $100 \ {\rm m}$, $\phi_s\in \{0^\circ, 120^\circ, 240^\circ\}$.}
  \label{fig:SimAll}
  \vspace{-0.5cm}
\end{figure*}

\begin{algorithm}[t!]
  \caption{Wavenumber-Domain Ellipse Fitting Algorithm}
  \label{WD-EF}
  \raggedright
  \textbf{Input:} $\widetilde{\mathcal{K}}_s$, $L_x$, $L_y$.

  \textbf{Output:} $\{r^x_s, r^y_s, r^z_s\}$, and $\{r_s, \theta_s, \phi_s\}$, $\forall s \in \mathcal{S}$.
  \vspace{-0.4cm}
  \begin{algorithmic}[1]
    \STATE Find the wavenumber-domain regions $\{\widetilde{\mathcal{K}}_s, s\in \mathcal{S}\}$.
    \FOR{each scatterer, i.e., $s\in \mathcal{S}$}
    \STATE {Extract the boundary sample set $\mathcal{B}^j_s$ from $\mathcal{K}_s$ by~\eqref{BoundaryExtraction}.}
    \STATE {Derive ellipse fitting results $\boldsymbol{\xi}^j_s$ by solving~\eqref{EllipseFitting}.}
    \STATE Obtain the estimated geometric parameters $\{r^x_s, r^y_s, r^z_s\}$ using~\eqref{rxEst}, \eqref{ryEst}, \eqref{rzEst}, respectively.
    \STATE Derive the spherical coordinates $\{r_s, \theta_s, \phi_s\}$ using~\eqref{est}.
    \ENDFOR
  \end{algorithmic}
\end{algorithm}

\section{Simulations}

\subsection{Simulation Setup}
The UPA is configured with $N_x = N_y = 512$ antennas with half-wavelength antenna spacing, working at the carrier frequency of $f_c = 7\ {\rm GHz}$.
Therefore, the antenna aperture is around $11 {\rm m}$, and the estimated Fresnel distance is $100\ {\rm m}$.
We deploy $S = 3$ scatterers within the Fresnel distance, each with uniformly spaced azimuth angles $\phi_s\in \{0^\circ, 120^\circ, 240^\circ\}$ and fixed elevation angles $\theta_s = 45^\circ$.
We then evaluate the normalized mean square error (NMSE) performance of the distance $({||[\hat{r}_s]_{s\in \mathcal{S}} - [{r}_s]_{s\in \mathcal{S}}||})/(|| [r_s]_{s\in \mathcal{S}} ||)$ and the angles of arrival (AoAs) $\frac{||[\hat{\theta}_s, \hat{\phi}_s]_{s\in \mathcal{S}} - [\theta_s, \phi_s]_{s\in \mathcal{S}}||}{||[\theta_s, \phi_s]_{s\in \mathcal{S}}||}$ under different distances and elevation angles, as shown in Fig.~\ref{fig:SimAll}(a) and Fig.~\ref{fig:SimAll}(b), respectively.
The benchmarks are as follows: 
\begin{itemize}
  \item \textbf{WD-EF}: The proposed scheme in \textbf{Algorithm~\ref{WD-EF}}.
  \item \textbf{GMM}: Approximating the observed wavenumber-domain spectrum region $\widetilde{\mathcal{K}}_s$ by a Gaussian mixed model (GMM)~\cite{guo-ICC24}, the AoAs are estimated by the statistical mean of the Gaussian distribution. 
  The distance is estimated by building a linear relationship between the Gaussian standard deviation $\sigma_x, \sigma_y$ and the distance, i.e., $ \hat{r}_s = \gamma \cdot {\rm norm}([\sigma_x, \sigma_y])$, where $\gamma$ is an experimentally determined constant.
  \item \textbf{Power-Weighted GMM}: Modified based on the GMM, the statistical center of the Gaussian distribution is weighted by the observed wavenumber-domain spectrum.
\end{itemize}  

\subsection{Simulation Evaluation}

\subsubsection{Robustness Beyond the Fresnel Bound}
We first conduct a performance evaluation across distances ranging $r_s \in [12.5, 110]$ (Fresnel distance is around $100\ {\rm m}$), with results depicted in Fig.~\ref{fig:SimAll}(a).
It is evident that traditional methods based on GMM or power-weighted GMM inherently perceive the shape in the wavenumber domain as elliptical. In the below-Fresnel region, the actual shape in the wavenumber domain deviates significantly from this assumption, leading to a rapid deterioration in performance as the distance decreases.
However, the proposed WD-EF algorithm directly utilizes the curve information from the wavenumber-domain spectrum. This means that as long as the boundaries, i.e., $\{\mathcal{B}^j_s, j\in \{1,\dots,4\}\}_{s\in \mathcal{S}}$ can be effectively extracted, the performance remains unaffected by distance. 
Consequently, the WD-EF algorithm maintains its effectiveness despite changes in distance within the below Fresnel region, and it becomes even more accurate as the distance shortens and the arcs in the wavenumber-domain spectrum lengthen.

\subsubsection{Robustness Over the Elevations}
In Fig.~\ref{fig:SimAll}(b), we plot the NMSE performance of benchmarks versus the elevation angle, i.e., $\theta_s \in [30^\circ,70^\circ]$.
It should be highlighted that although the GMM achieves accuracy in distance estimation comparable to the WD-EF algorithm presented in this paper, as shown in Fig.~\ref{fig:SimAll}(a), this accuracy does not hold across all elevation angles. This is because the accuracy of GMM's distance estimation heavily depends on the choice of the empirical parameter $\gamma$.
As demonstrated in Fig.~\ref{fig:SimAll}(b), once $\gamma$ is fixed, GMM's estimation performance peaks only within a specific range of elevation angles. Outside this range, performance deteriorates rapidly, with NMSE worsening to above $10^{-1}$ at angles less than $30^\circ$ or greater than $70^\circ$, which is unacceptable.

\section{Conclusion}
This paper investigates the wavenumber-domain channel estimation problem where the scatterers are located at the reactive near field below the Fresnel distance.
Specifically, we first derive closed-form expressions for the wavenumber-domain spectrum of spherical wavefronts emitted by scatterers. The boundaries of this observed spectrum are determined by ellipses, which in turn are defined by both the scatterers' geometric parameters and the array geometry.
Consequently, this paper introduces a novel WD-EF method, which precisely recovers the scatterers' Cartesian and spherical coordinates by leveraging the curve information within the elliptic spectrum boundaries.
Finally, simulation results demonstrate that in the reactive near-field region beneath the Fresnel distance, our proposed WD-EF algorithm exhibits significantly enhanced estimation accuracy and robustness when compared to existing benchmarks.

\section*{ACKNOWLEDGMENT}
This work was supported in part by the National Natural Science Foundation of China under Grant U22B2057, in part by the Beijing Natural Science Foundation under Grant 4222011, and in part by the BUPT Excellent Ph.D. Students Foundation under Grant CX2023145. (\textit{Corresponding authors: Ying Wang; Zhaocheng Wang.})

\bibliographystyle{IEEEtran}
\bibColoredItems{blue}{chen-twc} 
\bibColoredItems{blue}{guo-tcom} 
\bibliography{Holo_Spectrum}                       

\begin{thebibliography}{10}
\providecommand{\url}[1]{#1}
\csname url@samestyle\endcsname
\providecommand{\newblock}{\relax}
\providecommand{\bibinfo}[2]{#2}
\providecommand{\BIBentrySTDinterwordspacing}{\spaceskip=0pt\relax}
\providecommand{\BIBentryALTinterwordstretchfactor}{4}
\providecommand{\BIBentryALTinterwordspacing}{\spaceskip=\fontdimen2\font plus
\BIBentryALTinterwordstretchfactor\fontdimen3\font minus
  \fontdimen4\font\relax}
\providecommand{\BIBforeignlanguage}[2]{{%
\expandafter\ifx\csname l@#1\endcsname\relax
\typeout{** WARNING: IEEEtran.bst: No hyphenation pattern has been}%
\typeout{** loaded for the language `#1'. Using the pattern for}%
\typeout{** the default language instead.}%
\else
\language=\csname l@#1\endcsname
\fi
#2}}
\providecommand{\BIBdecl}{\relax}
\BIBdecl

\bibitem{HoloMag}
\BIBentryALTinterwordspacing
Y.~Chen, X.~Guo, G.~Zhou, S.~Jin, D.~W.~K. Ng, and Z.~Wang, ``Unified far-field
  and near-field in holographic {MIMO}: A wavenumber-domain perspective,''
  \emph{IEEE Commun. Mag.}, to appear, 2024. [Online]. Available:
  \url{https://arxiv.org/abs/2407.14815}
\BIBentrySTDinterwordspacing

\bibitem{chen}
Y.~Chen, Y.~Wang, Z.~Wang, and Z.~Han, ``Angular-distance based channel
  estimation for holographic {MIMO},'' \emph{IEEE J. Sel. Areas Commun.},
  vol.~42, no.~6, pp. 1684--1702, Jun. 2024.

\bibitem{TieruiGong}
T.~Gong, P.~Gavriilidis, R.~Ji, C.~Huang, G.~C. Alexandropoulos, L.~Wei,
  Z.~Zhang, M.~Debbah, H.~V. Poor, and C.~Yuen, ``Holographic {MIMO}
  communications: {Theoretical} foundations, enabling technologies, and future
  directions,'' \emph{IEEE Commun. Surv. Tutor.}, vol.~26, no.~1, pp. 196--257,
  First-quater, 2024.

\bibitem{DLL-ULA}
Z.~Wu and L.~Dai, ``Multiple access for near-field communications: {SDMA} or
  {LDMA}?'' \emph{IEEE J. Sel. Areas Commun.}, vol.~41, no.~6, pp. 1918--1935,
  Jun. 2023.

\bibitem{DLL-UPA}
Z.~Wu, M.~Cui, and L.~Dai, ``Multiple access for near-field communications:
  {SDMA} or {LDMA}?'' \emph{IEEE J. Sel. Areas Commun.}, vol.~41, no.~6, pp.
  1918--1935, Jun. 2023.

\bibitem{guo-ICC24}
X.~Guo, Y.~Chen, Y.~Wang, Z.~Wang, and Z.~Han, ``Wavenumber domain sparse
  channel estimation in holographic {MIMO},'' in \emph{IEEE International
  Conference on Communications}, Denver, CO, USA, Jun. 2024.

\bibitem{Guo-TPD}
X.~Guo, Y.~Chen, and Y.~Wang, ``Compressed channel estimation for near-field
  {XL-MIMO} using triple parametric decomposition,'' \emph{IEEE Trans. Veh.
  Technol.}, vol.~72, no.~11, pp. 15\,040--15\,045, Nov. 2023.

\bibitem{Yuanwei-Mag}
\BIBentryALTinterwordspacing
Y.~Liu, C.~Ouyang, Z.~Wang, J.~Xu, X.~Mu, and A.~L. Swindlehurst, ``Near-field
  communications: {A} comprehensive survey,'' \emph{arXiv:2401.05900}, Jan.
  2024. [Online]. Available: \url{https://arxiv.org/abs/2401.05900}
\BIBentrySTDinterwordspacing

\bibitem{Fourier}
A.~Pizzo, L.~Sanguinetti, and T.~L. Marzetta, ``Fourier plane-wave series
  expansion for holographic {MIMO} communications,'' \emph{IEEE Trans. Wireless
  Commun.}, vol.~21, no.~9, pp. 6890--6905, Sep. 2022.

\bibitem{SilhouetteCoefficient}
S.~Luan, X.~Kong, B.~Wang, Y.~Guo, and X.~You, ``Silhouette coefficient based
  approach on cell-phone classification for unknown source images,'' in
  \emph{IEEE International Conference on Communications}, Ottawa, Canada, Jun.
  2012, pp. 6744--6747.

\bibitem{Canny}
P.~Bao, L.~Zhang, and X.~Wu, ``Canny edge detection enhancement by scale
  multiplication,'' \emph{IEEE Trans. Pattern Anal. Mach. Intell.}, vol.~27,
  no.~9, pp. 1485--1490, Spe. 2005.

\bibitem{DirectLSEllipseFitting}
A.~Fitzgibbon, M.~Pilu, and R.~Fisher, ``Direct least square fitting of
  ellipses,'' \emph{IEEE Trans. Pattern Anal. Mach. Intell.}, vol.~21, no.~5,
  pp. 476--480, May 1999.

\end{thebibliography}

\end{document}